\documentclass[preprint,aps,prb,showpacs,amsmath,amssymb]{revtex4-1}

\usepackage{graphicx}			
\usepackage{bm}           
\usepackage{natbib}       

\usepackage{ifpdf}

\ifpdf
  \usepackage[pdftex]{hyperref}
  \hypersetup{
    colorlinks=false,
    pdfpagemode=UseThumbs,
    plainpages=false
   }
\else
  \usepackage[dvips]{hyperref}
  \hypersetup{ 
    colorlinks=false,
    pdfpagemode=UseThumbs,
    plainpages=false
   }
\fi

\begin{document}

\title[Electrostatically defined Quantum Dots in a Si/SiGe Heterostructure]{Electrostatically defined Quantum Dots in a Si/SiGe Heterostructure}

\author{A.~Wild$^1$, J.~Sailer$^1$, J.~N\"utzel$^1$, G.~Abstreiter$^1$, S.~Ludwig$^3$, D.~Bougeard$^{1,2}$}
\address{$^1$ Walter Schottky Institut, Technische Universit\"at M\"unchen, Am Coulombwall 3, 85748 Garching, Germany}
\address{$^2$ Institut f\"ur Experimentelle und Angewandte Physik, Universit\"at Regensburg, 93040 Regensburg, Germany}
\address{$^3$ Fakult\"at f\"ur Physik and Center for NanoScience, Ludwig-Maximilians-Universit\"at M\"unchen, Geschwister Scholl-Platz, 80539 M\"unchen, Germany}

\email{bougeard@wsi.tum.de}

\begin{abstract}
We present an electrostatically defined few-electron double quantum dot (QD) realized in a molecular beam epitaxy grown Si/SiGe heterostructure. Transport and charge spectroscopy with an additional QD as well as pulsed-gate measurements are demonstrated. We discuss technological challenges specific for silicon-based heterostructures and the effect of a comparably large effective electron mass on transport properties and tunability of the double QD. Charge noise, which might be intrinsically induced due to strain-engineering is proven not to affect the stable operation of our device as a spin qubit. Our results promise the suitability of electrostatically defined QDs in Si/SiGe heterostructures for quantum information processing. 
\end{abstract}

\pacs{73.23.Hk,73.63.Kv} 


\maketitle

\section{Introduction}
\label{sec:p01_Intro}

Electrostatically defined quantum dot (QD) structures are attracting increasing interest as building blocks for solid state based quantum information processing. In such structures, the electron spin decoherence time is crucial for coherent manipulation of spin qubits. Electron spin phenomena have already been investigated in QD structures in AlGaAs/GaAs heterostructures~\cite{Hanson2007}. The hyperfine interaction of the electrons confined in such QDs with roughly $\mathrm{10^5}$ thermally fluctuating nuclear spins has been identified as a limiting decoherence mechanism for electron spin qubits in GaAs~\cite{Merkulov2002,Witzel2006,Coish2008}. This problem can be addressed by manipulating nuclear spins in GaAs \cite{Petta2008,Vink2009,Foletti2009} or by choosing an alternative host material. Silicon (Si) as a host material, offers a promising path towards extending the electron spin coherence time compared to GaAs based qubits, because naturally composed Si-crystals contain only a fraction of about $4.7\%$ of nuclear spin carrying isotopes~\cite{Ager2006} compared to $\mathrm{100\%}$ in GaAs. Since the hyperfine interaction strength is roughly proportional to the fraction of nuclear spin carrying isotopes, much longer coherence times are expected for Si. Furthermore, Si has a weaker spin-orbit interaction~\cite{Tahan2005,Prada2010} and is not piezo-electric~\cite{Prada2008}.  

Recently promising efforts towards the implementation of spin qubits in natural Si have been made in three classes of devices: electrostatically defined QDs in strain engineered Si/SiGe heterostructures~\cite{Simmons2007,Shaji2008,Simmons2009,Hayes2009,Thalakulam2010}, QDs in MOS structures~\cite{Liu2008,Lim2009,Nordberg2009,Hu2009,Xiao2010} or MOS-phosphorus hybrid devices~\cite{Prati2009,Huebl2009,Morello2009,Tan2010,Morello2010}. In some of these devices, first pulsed-gate experiments on Si-based QDs have been performed, including spin relaxation time measurements ($T_1$) of a confined electron~\cite{Hayes2009,Xiao2010,Morello2010}, the demonstration of single-shot readout~\cite{Morello2010} and the measurement of tunneling rates~\cite{Huebl2009,Thalakulam2010}.

In this emerging research field, open questions remain, such as the influence of Si/SiGe specific material properties on device performance and tunablility. In this contribution, we present a Si/SiGe heterostructure whose material properties can be precisely controlled in molecular beam epitaxy (MBE). The heterostructure contains a strain-induced high mobility two-dimensional electron system (2DES) and is equipped with metallic top gates. In the resulting device, we implement a double QD combined with a single electron transistor (SET) as a charge sensor, both tunable by the field effect. An important fundamental difference of Si- to GaAs-based structures is the considerably larger effective electron mass in the 2DES ($m_{\mathrm{e,Si}}^* =0.19\cdot m_{\mathrm{e}}\approx \mathrm{3}\cdot m_{\mathrm{e,GaAs}}^*$). We discuss the direct consequences of a high electron mass which can be observed \emph{e.g.} in form of a small Fermi-energy of the 2DES and low tunneling rates of electrons across electrostatic barriers. 
In our measurements, charge noise strongly affects a large scale stability diagram, but is a minor issue as long as gate voltages are changed only slightly. Thereby, we demonstrate stable operation of our double QD device and its suitability as a spin qubit.

\section{Material and Sample Development}
\label{sec:p02_MaterialSample}

\begin{figure}
\begin{center}
\includegraphics[width=0.6\columnwidth]{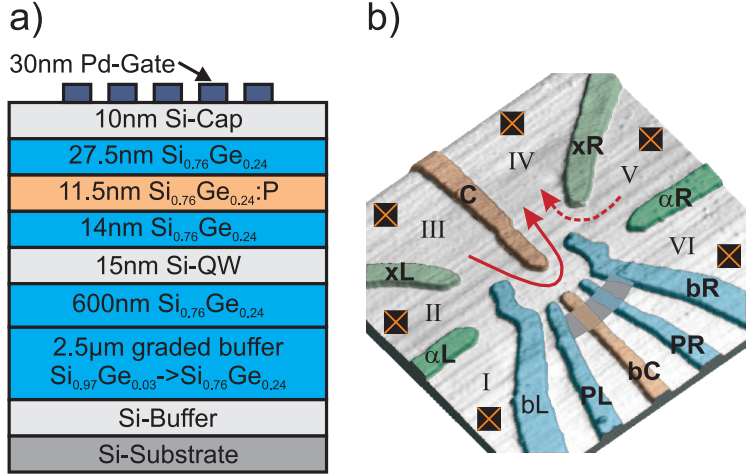}
\end{center}

\caption{a) Layout of the MBE Si/SiGe heterostructure showing the layer structure. Evaporated palladium gates deplete parts of the two-dimensional electron gas in the silicon quantum well (Si-QW). b) AFM micrograph of the gates defining the double QD. Due to a short-circuit symbolized by a bold stripe, gates PL, bC, PR and bR are on an identical potential. The occupation of the double QD is tuned via gate voltages $V_{\mathrm{bL}}$ and $V_{\mathrm{bR}}$. In a direct transport spectroscopy experiment, current flows between contacts III and IV (solid arrow). In addition, a single QD, defined by gate bR and xR, can be used as a charge sensor. In this case current flows between the contacts V and IV (dashed arrow).}

\label{fig:Fig1}
\end{figure}

Our double QD is electrostatically formed within a 2DES in a strained-Si quantum well (QW) of a MBE grown Si/Si$_{\mathrm{1}-x}$Ge$_{x}$ heterostructure with $x=\mathrm{24~\%}$. The heterostructure layout and composition is shown in figure~\ref{fig:Fig1}(a). A layer doped by phosphorus gives rise to a maxium 2DES density of about $\mathrm{3.5\times10^{11}~cm^{-2}}$ and an electron mobility of $\mathrm{1.1\times10^5~cm^{2}\,(Vs)^{-1}}$ in this wafer at the temperature $T=\mathrm{1.4~K}$. The biaxial tensile strain in the Si QW lifts the sixfold valley degeneracy of bulk Si. The energy of the two valleys in [0~0~1] growth direction is lowered by $\mathrm{230~meV}$ below the conduction band edge of the surrounding Si$\mathrm{_{0.76}}$Ge$\mathrm{_{0.24}}$ layers. From a one-dimensional self-consistent band structure calculation with next\textbf{nano}++~\cite{Trellakis2006,Birner2007}, we obtain an intravalley subbband spacing between the first two subbands on the order of $\mathrm{8~meV}$ which is large compared to the Fermi energy of $E_{\mathrm{F}}=\mathrm{1.1~meV}$. This small Fermi-energy, compared to typical GaAs heterostructures, is a consequence of the high effective electron mass in Si and the two-fold valley degeneracy.

The double QD is defined in a mesa fabricated by wet-chemical etching. Ohmic contacts are formed by diffusing Sb/Au into the heterostructure. Electric top-gates are fabricated by electron beam lithography and palladium (Pd) evaporation~\cite{Sailer2009}. The Pd on the device surface pins the Fermi energy at about $\mathrm{750~meV}$~\cite{Kircher1971,Tongson1979} below the conduction band edge. This strong pinning is as consequence of surface states at the Pd-Si interface. The surface states bind most of the electrons otherwise remaining at the doping layer. Together with the high work function of Pd, this results in a large Schottky barrier. The latter helps to minimize leakage currents from biased gates into the heterostructure. Our double QD gate design has been adapted from comparable GaAs-based structures~\cite{Elzerman2003}. A nominally identical device to the one investigated in this work is shown in an AFM micrograph in figure~\ref{fig:Fig1}(b). High frequency coaxial cables lead to the gates bL and bR on the sample surface while all other gates are connected via low-frequency wires. After cool-down to $T_{\mathrm{2DES}} \approx \mathrm{100~mK}$, samples from the studied wafer require weak illumination with a red LED in order to populate the 2DES. We find that even at zero applied bias, the mere presence of Pd on top of the Si cap layer completely depletes the 2DES underneath. This behavior has been observed before~\cite{Holzmann1994,Tobben1995a,Tobben1995b} and is mainly a consequence of the saturation of Si dangling bonds~\cite{Tao2003} and the related Fermi level pinning at the Pd-Si interface~\cite{Schaffler1997} at low doping concentrations. Consequently, positive voltages are typically applied to all gates in order to drive currents from ohmic contacts III or V to IV.
An unintended electrical short between gates PL, bC, PR and bR forces these gates to be on the same electrical potential. We will refer to this potential as $V_{\mathrm{bR}}$ in the following. As a consequence of the short, the inter-dot barrier, the energy levels and the tunnel barriers from both dots to the leads cannot be tuned independently.

\section{Results and Discussion}

\subsection{Transport Spectroscopy}
\label{sec:p03_Transport}

\begin{figure}
\begin{center}
\includegraphics[width=0.6\columnwidth]{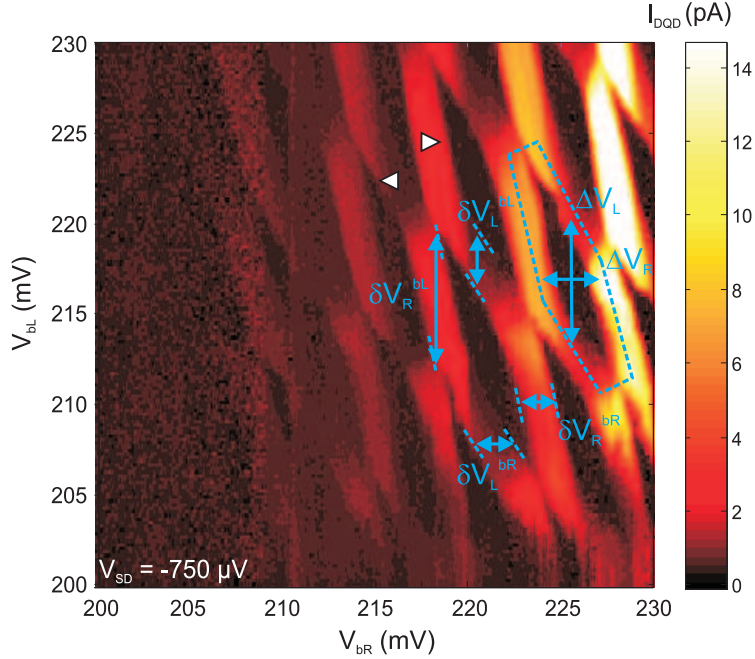}
\end{center}

\caption{Charge stability diagram which shows the dc current $I_{\mathrm{DQD}}$ as a function of $V_{\mathrm{bL}}$ and $V_{\mathrm{bR}}$ flowing through the double QD, while a bias of $\mathrm{-750~\mu V}$ is applied at contact III. The width of the charging lines $\delta V_{\mathrm{L}}^{\mathrm{bL,bR}}$ and $\delta V_{\mathrm{R}}^{\mathrm{bL,bR}}$ correspond to an energy window (transport window) of $eV_{\mathrm{SD}}=\mathrm{750~\mu eV}$ defined by the applied bias. They define the lever arms $\alpha_{\mathrm{L}}^{\mathrm{bL}} \approx \mathrm{0.18~e}$, $\alpha_{\mathrm{R}}^{\mathrm{bR}} \approx \mathrm{0.39~e}$, $\alpha_{\mathrm{L}}^{\mathrm{bR}} \approx \mathrm{0.31~e}$ and $\alpha_{\mathrm{R}}^{\mathrm{bL}} \approx \mathrm{0.07~e}$ and allow a conversion of the voltages $V_{\mathrm{bL}}$ and $V_{\mathrm{bR}}$ to energy. The distances between consecutive charging lines $\Delta V_{\mathrm{L}}$ and $\Delta V_{\mathrm{R}}$ can be converted to the charging energies $E_{\mathrm{CL}}\approx\mathrm{1.5~meV}$ and  $E_{\mathrm{CR}}\approx\mathrm{1.6~meV}$ of the QDs.}

\label{fig:Fig2}
\end{figure}

Figure~\ref{fig:Fig2} depicts the dc current $I_{\mathrm{DQD}}$ flowing across the serial double QD as a function of the gate voltages  $V_{\mathrm{bL}}$ and $V_{\mathrm{bR}}$. $I_{\mathrm{DQD}}$ is  measured at contact IV while a dc bias of $V_{\mathrm{SD}}=\mathrm{-750~\mu V}$ is applied to contact III. The charge stability diagram shows the typical honeycomb pattern reminiscent of a double QD. $I_{\mathrm{DQD}}>\mathrm{0}$ is observed on the triple points, but also along charging lines. $I_{\mathrm{DQD}}$ completely disappears below $V_{\mathrm{bR}}\leq\mathrm{205~mV}$ and towards the lower left corner of the stability diagram. 
On the triple points, the charge fluctuates on both QDs and linear response transport by sequential tunneling is expected. However, in contrary to our observation, no current is expected along the charging lines as one of the QDs is in Coulomb blockade. In this regime, the observed non-zero $I_{\mathrm{DQD}}$ along charging lines is maintained by elastic and inelastic co-tunneling and enhanced by the comparatively large $V_{\mathrm{SD}}$. 

The widths $\delta V_{\mathrm{L}}^{\mathrm{bL,bR}}$ and $\delta V_{\mathrm{R}}^{\mathrm{bL,bR}}$ of the charging lines in figure~\ref{fig:Fig2} with respect to the gate voltage axis $V_{\mathrm{bL}}$ and $V_{\mathrm{bR}}$ serve as a calibration scale to convert the applied gate voltage into energy. We find the lever arms $\alpha_{\mathrm{L}}^{\mathrm{bL}}=e V_{\mathrm{SD}} / \delta V_{\mathrm{L}}^{\mathrm{bL}} \approx \mathrm{0.18~e}$, $\alpha_{\mathrm{R}}^{\mathrm{bR}}=e V_{\mathrm{SD}} / \delta V_{\mathrm{R}}^{\mathrm{bR}} \approx \mathrm{0.39~e}$, $\alpha_{\mathrm{L}}^{\mathrm{bR}}=e V_{\mathrm{SD}} / \delta V_{\mathrm{L}}^{\mathrm{bR}} \approx \mathrm{0.31~e}$ and $\alpha_{\mathrm{R}}^{\mathrm{bL}}=e V_{\mathrm{SD}} / \delta V_{\mathrm{R}}^{\mathrm{bL}} \approx \mathrm{0.07~e}$. The lever arms are then used to deduce typical charging energies of both QDs from the distances $\Delta V_{\mathrm{L}}$ and $\Delta V_{\mathrm{R}}$ between parallel charging lines in figure~\ref{fig:Fig2}. The charging energies are on the order of $E_{\mathrm{CL}}=\alpha_{\mathrm{L}}^{\mathrm{bL}} \cdot \Delta V_{\mathrm{L}} \leq \mathrm{1.5~meV}$ and $E_{\mathrm{CR}} = \alpha_{\mathrm{R}}^{\mathrm{bR}} \cdot \Delta V_{\mathrm{R}} \leq \mathrm{1.6~meV}$. Based on a 3D self-consistent band structure calculation performed with next\textbf{nano}++, where we take into account the dot capacitances for the given gate geometry, we estimate the double QD occupation of about $N_{\mathrm{L}} \approx N_{\mathrm{R}} \approx \mathrm{18}$ electrons\footnote{A simple disc model for the double QD based on the 2DES density and charging energies typically overestimates the occupation by a factor of 2 compared to the self-consistent band structure simulation.}.

The disappearance of $I_{\mathrm{DQD}}$ below $V_{\mathrm{bR}}\leq\mathrm{205~mV}$ and the overall large effective resistance $V_{\mathrm{SD}}/I_{\mathrm{DQD}}\geq\mathrm{60~M\Omega}$ for the stability diagram is in part caused by the large effective electron mass $m^*$ in Si-based 2DES since tunneling rates are exponentially suppressed as the mass of the tunneling particle is increased. However, in our device, the low $I_{\mathrm{DQD}}$ is furthermore a consequence of the short between gates PL, bC, PR and bR. This short not only results in a strong capacitive coupling of $V_{\mathrm{bR}}$ to the right, but also to the left QD and in a strong suppressing effect of $V_{\mathrm{bR}}$ on the QD-lead tunneling rates. Furthermore, $V_{\mathrm{bR}}$ can be expected to asymmetrically influence the tunneling rates of the left and right QD to its leads. 
We can observe the effect of asymmetric QD-lead tunneling rates in figure~\ref{fig:Fig2} in a larger current value along the charging lines of the right QD (marked by $\rhd$ in figure~\ref{fig:Fig2}) compared to the charging lines of the left QD (marked by $\lhd$ in figure~\ref{fig:Fig2}). Away from the triple points, current along the charging lines of the right (left) QD involves a first order tunneling process and a second order co-tunneling process in series. The current is roughly given by
\begin{equation*}
I_{\mathrm{DQD}}^{\rhd} \propto \frac{\Gamma_{\mathrm{L}} \Gamma_{\mathrm{iD}} \Gamma_{\mathrm{R}}}{\Gamma_{\mathrm{L}} \Gamma_{\mathrm{iD}} + \Gamma_{\mathrm{R}} \Delta/\hbar} 
\mathrm{~and~}  
I_{\mathrm{DQD}}^{\lhd} \propto \frac{\Gamma_{\mathrm{L}} \Gamma_{\mathrm{iD}} \Gamma_{\mathrm{R}}}{\Gamma_{\mathrm{R}} \Gamma_{\mathrm{iD}} + \Gamma_{\mathrm{L}} \Delta/\hbar} \mathrm{.} 
\label{eq:RateEquation}
\end{equation*}
Here, $\Gamma_{\mathrm{L}}$ and $\Gamma_{\mathrm{R}}$ are the respective QD-lead tunneling rates and $\Gamma_{\mathrm{iD}}$ is the inter-dot tunneling rate. The asymmetry energy $\Delta$ separates the energies of the localized states with the electron being either in the left or in the right QD. As we observe $I_{\mathrm{DQD}}^{\rhd} > I_{\mathrm{DQD}}^{\lhd}$ (where we use the realistic assumption $\Delta < \hbar\Gamma_{\mathrm{iD}}$), which corresponds to $\Gamma_{\mathrm{R}} < \Gamma_{\mathrm{L}}$, the tunnel coupling between the right QD and its lead is weaker than the tunneling coupling between the left QD and its lead.

\begin{figure}
\begin{center}
\includegraphics[width=0.6\columnwidth]{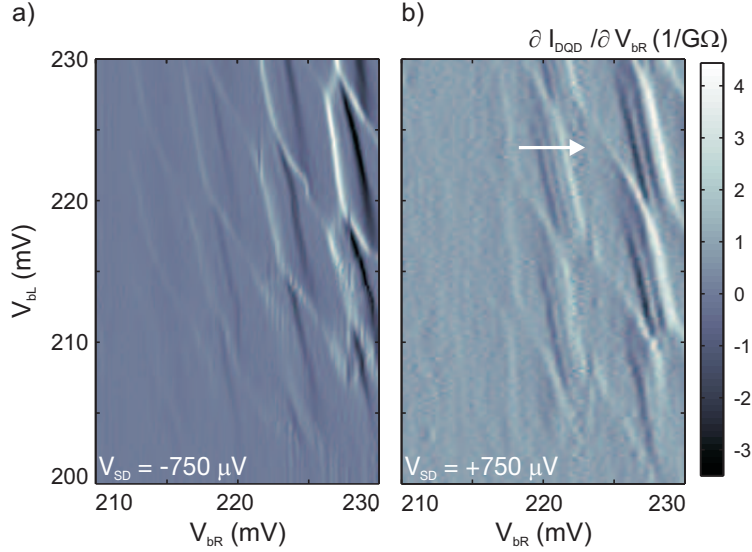}
\end{center}

\caption{Transconductance $\partial I_{\mathrm{DQD}}/\partial V_{\mathrm{bR}}$ of the double QD as a function of $V_{\mathrm{bL}}$ and $V_{\mathrm{bR}}$ for an applied bias of $\mathrm{-750~\mu V}$ (a) and $\mathrm{+750~\mu V}$ (b). Charging lines of the left QD are barely visible, whereas charging lines of the right QD exhibit an additional substructure only for $\mathrm{+750~\mu V}$. This substructure is not present for $\mathrm{-750~\mu V}$.}

\label{fig:Fig3}
\end{figure}

Figure~\ref{fig:Fig3} contrasts the transconductance $\partial I_{\mathrm{DQD}}/\partial V_{\mathrm{bR}}$ for an applied dc bias of $\mathrm{-750~\mu V}$, as in figure~\ref{fig:Fig2}, with the corresponding measurement for an applied bias of $\mathrm{+750~\mu V}$. Due to $\Gamma_{\mathrm{R}}<\Gamma_{\mathrm{L}}$, and in agreement with the current measurement in figure~\ref{fig:Fig2}, the charging lines of the right QD are more pronounced compared to those of the left QD. For positive $V_{\mathrm{SD}}$, a substructure within the charging line of the right QD of width $eV_{\mathrm{SD}} / \alpha_{\mathrm{R}}^{\mathrm{bR}}$ is visible in the form of parallel lines of alternating high and low transconductance. No such lines are observable for negative $V_{\mathrm{SD}}$.

\begin{figure}
\begin{center}
\includegraphics[width=0.6\columnwidth]{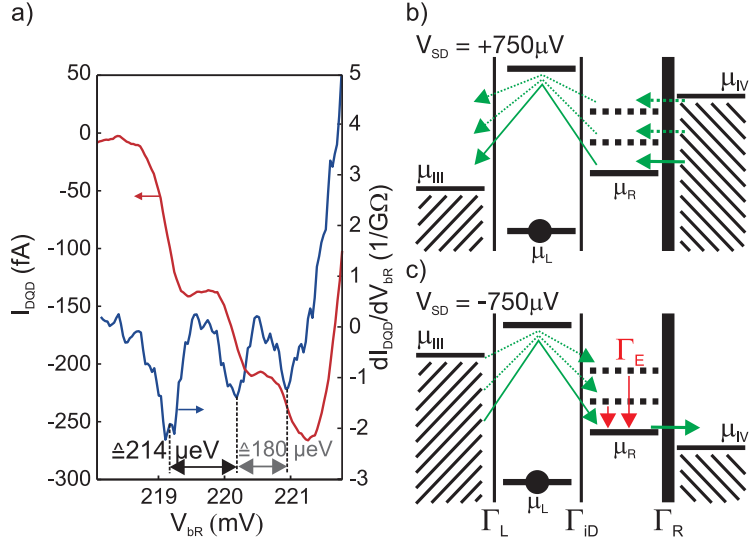}
\end{center}

\caption{a) Cross-section along the white arrow in figure~\ref{fig:Fig3}b). The double QD current $I_{\mathrm{DQD}}$ is shown on the left axis and its derivative, the transconductance $\partial I_{\mathrm{DQD}}/\partial V_{\mathrm{bR}}$, is shown on the right axis. Energy level configuration and likely transport channels (arrows) are shown for $V_{\mathrm{SD}}=\mathrm{+750~\mu V}$ in b) and $V_{\mathrm{SD}}=\mathrm{-750~\mu V}$ in c). Transport in the double QD takes place by resonant tunneling through the right QD and co-tunneling through the Coulomb-blocked left QD. }

\label{fig:Fig4}
\end{figure}

This situation is discussed in figure~\ref{fig:Fig4} in more detail. Figure~\ref{fig:Fig4}(a) plots $I_{\mathrm{DQD}}(V_{\mathrm{bR}})$ as well as the corresponding transconductance $\partial I_{\mathrm{DQD}}/\partial V_{\mathrm{bR}}\,(V_{\mathrm{bR}})$ for $V_{\mathrm{SD}}=\mathrm{+750~\mu V}$ along the horizontal white arrow in figure~\ref{fig:Fig3}(b). $|I_{\mathrm{DQD}}|$ increases in steps as $V_{\mathrm{bR}}$ is increased, while the transconductance shows corresponding oscillations. These observations are explained in figure~\ref{fig:Fig4}(b) by explicitly taking resonant tunneling  and co-tunneling processes into account. Electrons can tunnel resonantly from the right lead IV into the right QD followed by an elastic co-tunneling process via the Coulomb-blocked left QD into the left lead III. Note that inelastic co-tunneling processes are also possible, but do not change our qualitative argument. Dotted lines in figure~\ref{fig:Fig4}(b) mark the single particle excitation spectrum of the right QD. The observed current steps and transconductance oscillations in figure~\ref{fig:Fig4}(a) for $V_{\mathrm{SD}}>0$ imply that the excited states of the right QD contribute separately to $I_{\mathrm{DQD}}$ as depicted by arrows in figure~\ref{fig:Fig4}(b). This also implies that the energy relaxation rate $\Gamma_{\mathrm{E}}$ within the right QD is slow compared to the co-tunneling rates between the right QD and the left lead III. In figure~\ref{fig:Fig4}(a), we resolve two excited states with a characteristic excitation energy of approximately $\mathrm{200~\mu eV}$. 
For $V_{\mathrm{SD}}<0$, no excited states are observed along the charging lines of the right QD in figure~\ref{fig:Fig3}(a). Here a co-tunneling process is followed by resonant tunneling from the right QD to the right lead IV as sketched in figure~\ref{fig:Fig4}(c). However, energy relaxation in the right QD is fast compared to the slow tunneling rate $\Gamma_{\mathrm{R}}$. Hence, the missing excitation spectrum of the right QD for $V_{\mathrm{SD}}<0$ not only confirms the previous finding $\Gamma_{\mathrm{R}}<\Gamma_{\mathrm{L}}$, but furthermore suggests $\Gamma_{\mathrm{R}}\ll\Gamma_{\mathrm{L}}\mathrm{, }\Gamma_{\mathrm{iD}}$.

\subsection{Charge Sensing}
\label{sec:p04_ChargeSensing}

\begin{figure}
\begin{center}
\includegraphics[width=0.6\columnwidth]{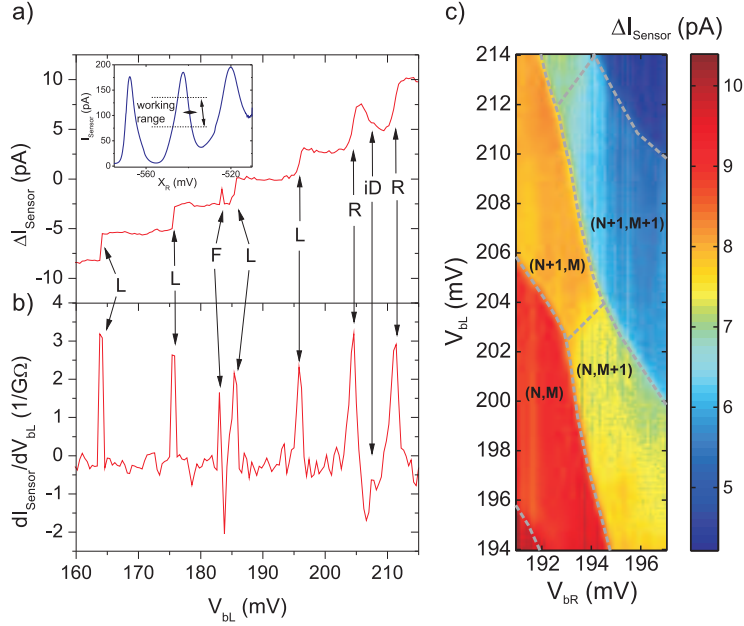}
\end{center}

\caption{ A QD single electron transistor as a charge sensor: a) The inset plots $I_{\mathrm{Sensor}}(V_{\mathrm{xR}})$, showing Coulomb blockade oscillations. A dedicated working range is marked. The main plot shows the sensor current after subtraction of a linear background $\Delta I_{\mathrm{Sensor}} = I_{\mathrm{Sensor}} - \gamma V_{\mathrm{bL}}$ for $V_{\mathrm{bR}}=\mathrm{200~mV}$. b) Numerical derivative $\partial \left( \Delta I_{\mathrm{Sensor}} \right) / \partial V_{\mathrm{bL}}$. c) Charge stability diagram in the vicinity of two triple points that shows $\Delta I_{\mathrm{Sensor}}$ as a function of $V_{\mathrm{bL}}$ and $V_{\mathrm{bR}}$ for different stable ground state charge configurations (different colors).}

\label{fig:Fig5}
\end{figure}

In an attempt to characterize the double QD in the few-electron regime, we use charge sensing~\cite{Elzerman2003} via a QD SET. The QD is located between gates bR, $\mathrm{\alpha_R}$ and xR. Current is measured from contact V to IV as illustrated by the dashed arrow in figure~\ref{fig:Fig1}(b). The Coulomb-blockade oscillations of the sensor QD are plotted in the inset of figure~\ref{fig:Fig5}(a) for a sensor bias of $V_{\mathrm{SD}}=\mathrm{400~\mu eV}$. Its charging energy is approximately $E_{\mathrm{C}} \approx \mathrm{1.5~meV}$. As the working range of the charge sensor, we choose one of the flanks of a Coulomb-peak were $|\partial I_{\mathrm{Sensor}}/\partial V_{\mathrm{xR}}|$ and thus the sensor sensitivity have local maxima. The main plot of figure~\ref{fig:Fig5}(a) shows $\Delta I_{\mathrm{Sensor}} = I_{\mathrm{Sensor}} - \gamma V_{\mathrm{bL}}$, that is the sensor current after subtraction of a straight line ($\gamma V_{\mathrm{bL}}$). It represents the direct capacitive coupling between the sensor QD and gate bL as a function of $V_{\mathrm{bL}}$. The pronounced steps mark single electron charging events of the double QD. 
The transconductance $\partial \left( \Delta I_{\mathrm{Sensor}} \right) / \partial V_{\mathrm{bL}}$ plotted in figure~\ref{fig:Fig5}(b) exhibits sharp local maxima where $I_{\mathrm{Sensor}}$ has steps. The charge sensor has a resolution of at least 0.1 electron charges determined from its signal-to-noise ratio. More importantly, it allows to measure the charge stability diagram in a regime where a current through the double QD is already too small to be detected by standard techniques. This is especially important in order to control the few-electron regime in Si-based serially coupled QDs where the large effective mass of the electrons causes weak tunnel couplings. 

From the step heights in $\Delta I_{\mathrm{Sensor}}$ and the peak shape in  $\partial I_{\mathrm{Sensor}} / \partial V_{\mathrm{bL}}$, we can distinguish charging events of the left (L) or right (R) QD, a nearby charge trap (F) or inter-dot (iD) transitions. Since the capacitive coupling between the right QD and the sensor is stronger, the associated steps are higher (R) compared to steps associated with charging events of the left QD (L). An inter-dot transition (iD) where an electron moves from the right to the left QD results in a decrease of $\Delta I_{\mathrm{Sensor}}$ and thus in a transconductance minimum. In addition, a nearby charge fluctuation causes a local maximum of $\Delta I_{\mathrm{Sensor}}$ (F).

During long measurements over a wide plunger gate voltage range, cross-talk and charge fluctuations in the environment can cause a considerable drift of the sensor working point out of its working range which we define by horizontal lines in the inset of figure~\ref{fig:Fig5}(a). In order to avoid such a drift of the sensor, gate xR is used for stabilization of the local potential of the charge sensor. We perform a linear adjustment of $V_{\mathrm{xR}}(V_{\mathrm{bL}})$ during each plunger gate sweep and a stepwise adjustment after each $V_{\mathrm{bR}}$ step. 
This technique has been used for the charge stability diagram in figure~\ref{fig:Fig5}(c). It plots the sensor current as function of $V_{\mathrm{bL}}$ and $V_{\mathrm{bR}}$ from which a plane fit has been subtracted. Regions of stable ground state charge configurations are marked by $(N,M)\leftrightarrow(N+\mathrm{1},M+\mathrm{1})$ for $N$, $N+\mathrm{1}$ electrons in the left QD and $M$, $M+\mathrm{1}$ electrons in the right QD. 

\begin{figure}
\begin{center}
\includegraphics[width=0.6\columnwidth]{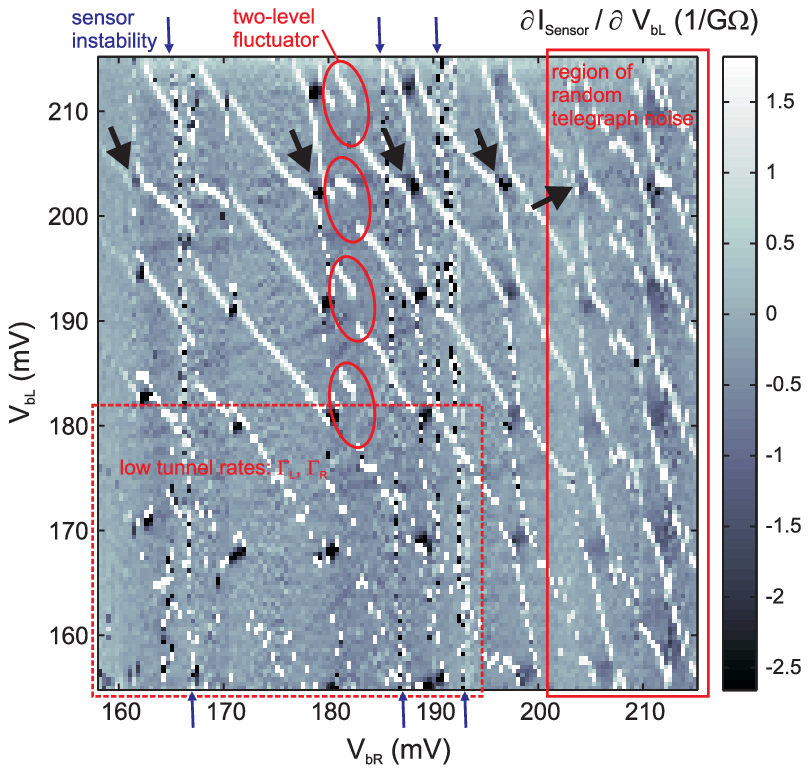}
\end{center}

\caption{ Large scale charge stability diagram showing the sensor transconductance $\partial I_{\mathrm{Sensor}}/\partial V_{\mathrm{bL}}$. Stable regimes alternate with noisy areas. Charge noise which mainly affects the sensor QD is marked by blue arrows. A specific fluctuator which causes a bistability is marked by red ellipses and a region of overall strong charge noise which directly affects the double QD is framed by a red box. The lower left corner (dashed  box) features very low QD-lead tunneling rates $\Gamma_{\mathrm{L}}$ and $\Gamma_{\mathrm{R}}$.}

\label{fig:Fig6a}
\end{figure}

Figure~\ref{fig:Fig6a} shows the transconductance $\partial I_{\mathrm{Sensor}}/\partial V_{\mathrm{bL}}$ for a larger gate voltage regime compared to figure~\ref{fig:Fig5}(c). Charging lines are white ($\partial I_{\mathrm{Sensor}} / \partial V_{\mathrm{bL}} > \mathrm{0}$) whereas reconfiguration lines are black ($\partial I_{\mathrm{Sensor}} / \partial V_{\mathrm{bL}} < \mathrm{0}$). This large scale stability diagram was obtained by sweeping $V_{\mathrm{bL}}$ from $\mathrm{220~mV}$ to $\mathrm{150~mV}$ and stepping $V_{\mathrm{bR}}$ from high to low voltages. The plot contains stable regions, but is riddled with noisy areas. Most of the observed noise can be attributed to fluctuating charges in the vicinity of the nanostructure (charge noise). For $V_{\mathrm{bR}}>\mathrm{200~mV}$, local potential fluctuations at the double QD are present and result in random telegraph noise. In contrast, for $V_{\mathrm{bR}}<\mathrm{200~mV}$ the local potential of the double QD is relatively stable. The red ellipses at $V_{\mathrm{bR}} \approx \mathrm{182~mV}$ mark a bistability caused by a specific two-level charge fluctuator with a characteristic time constant of several minutes. Fluctuations running almost vertically through the diagram are marked by blue arrows at the top and bottom of the plot. They do hardly affect the charging lines, but primarily the background signal at stable double QD charge configurations. They can be attributed to local potential fluctuations restricted to the sensor QD.

The stability diagram of figure~\ref{fig:Fig6a} shows the general tendency that the appearance of fluctuations strongly depends on $V_{\mathrm{bR}}$ while $V_{\mathrm{bL}}$ has almost no influence. This can be interpreted as a hint that the observed telegraph noise is not a general problem of the heterostructure, but is rather linked to the unintended short between the gates PL, bC, PR and bR, all lying on the same potential. Lateral leakage currents along the sample surface are likely to cause the short. These leakage currents can also trigger charge fluctuations which result in the observed telegraph noise. 

In addition, the risk of vertical leakage currents is higher for strain-engineered Si/SiGe heterostructures compared to AlGaAs/GaAs based structures. The plastic strain relaxation process during SiGe epitaxy leads to the formation of threading dislocations~\cite{Beanland1996,Mooney1996} (TD) which can pierce all the way from the graded buffer SiGe layer [see figure~\ref{fig:Fig1}(a)] through the strained Si quantum well to the surface. 
These crystal defects in Si/SiGe heterostructures are associated with mid-band gap states along the TD. The latter are a possible source of charge fluctuations~\cite{Mooney2000,Berbezier2002} at the surface. Furthermore, the presence of TDs could facilitate vertical leakage currents from a biased gate into the heterostructure. However, in the case of figure~\ref{fig:Fig6a}, such leakage currents are too small to be directly observed. Transmission electron microscopy and atomic force microscopy measurements of our heterostructures, suggest TD densities typically lower than $\mathrm{10^7~cm^{-2}}$. This corresponds to a probability of less than $\mathrm{5~\%}$ to find even a single TD in the direct vicinity of the double QD or the sensor QD. Hence, TDs are expected to play only a minor role to the charge noise observed in figure~\ref{fig:Fig6a}.

In general, our charge sensing experiments demonstrate an important tendency. Despite of locally strong charge noise for voltage sweeps covering large a range as in figure~\ref{fig:Fig6a}, stable operation of the double QD, as well as the sensor, is possible in small gate voltage intervals. Irrespective of the exact origin of occasional charge noise, this local stability is more essential and should allow the successful realization and stable operation of qubits.

\begin{figure}
\begin{center}
\includegraphics[width=0.9\columnwidth]{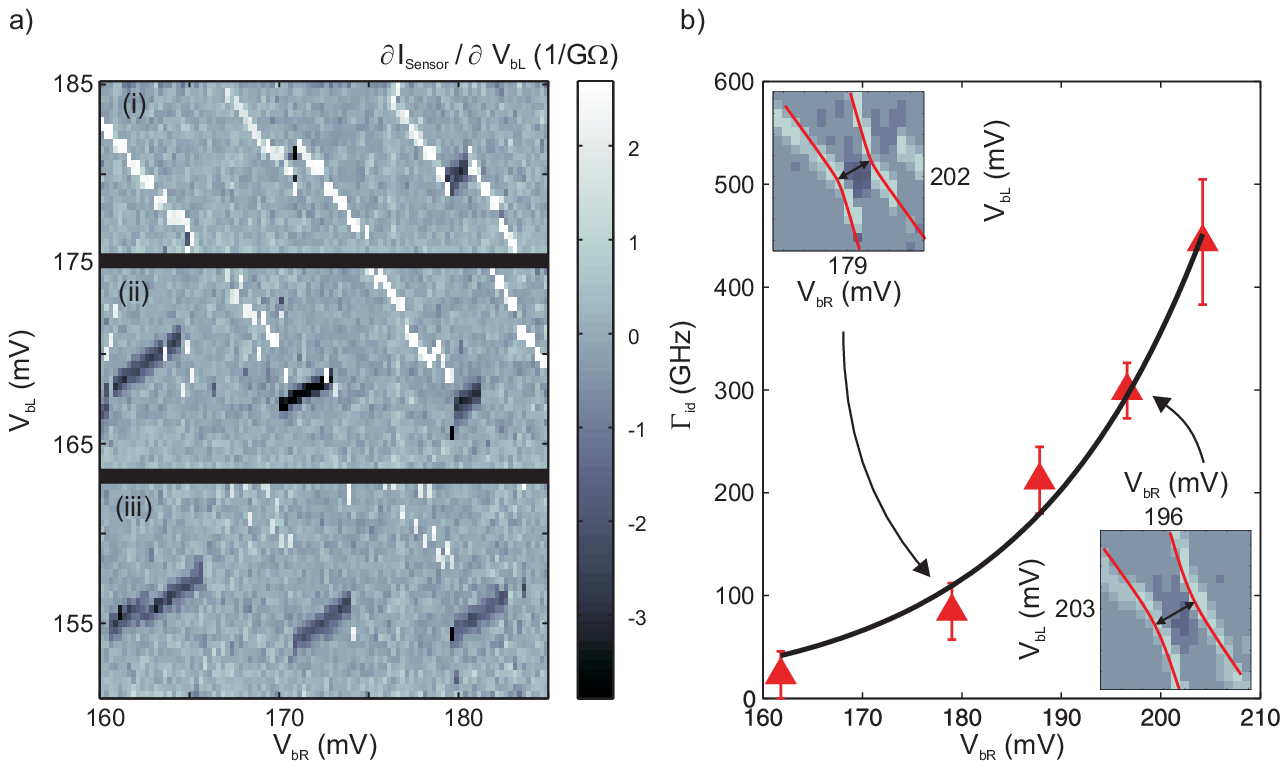}
\end{center}

\caption{a) Charge stability diagram in the lower left isolation regime of figure~\ref{fig:Fig6a}. The diagram is composed of three sections (i), (ii) and (iii) that have been measured successively. The tunneling rates change from $\Gamma >> \mathrm{1~Hz}$ to $\Gamma << \mathrm{1~Hz}$ in the depicted regime, giving rise to discontinuous or absent charging lines. b) Inter-dot tunneling rate $\Gamma_{\mathrm{id}}(V_{\mathrm{bR}})$. Triangles are obtained by fitting avoided crossings near triple points marked by black arrows in figure~\ref{fig:Fig6a}. Insets show exemplary avoided crossings in the stability diagram. The solid line in the main plot is a WKB-fit.}

\label{fig:Fig6b}
\end{figure}

We now turn to the lower left part of figure~\ref{fig:Fig6a}, encircled with a dashed box. Here, the number of electrons charging each QD is well below 10. Below $V_{\mathrm{bL}}=\mathrm{180~mV}$ and for decreasing values of $V_{\mathrm{bR}}$, first, charging lines associated with the right QD and then, those associated with the left QD, become discontinuous or even vanish. In contrast to the previous discussion, this phenomenon is not caused by charge noise, but results from decreasing tunnel couplings between the QDs and their leads as the plunger gate voltages are decremented~\cite{Rushforth2004,Thalakulam2010}.
More accurate measurements of the dc sensor current $I_{\mathrm{Sensor}}$ in such a regime are shown in figure~\ref{fig:Fig6b}(a). The data was taken by sweeping $V_{\mathrm{bL}}$ at a rate of $\mathrm{0.5~mV\,s^{-1}}$ (vertical sweeps from top to bottom of section I take $t_{\mathrm{sweep}}=\mathrm{10~s}$) and stepping $V_{\mathrm{bR}}$ from the right to the left in between vertical sweeps. The data was measured subsequently for the three intervals (i), (ii) and (iii). Here, we plot the numerical derivative $\partial I_{\mathrm{Sensor}}/\partial V_{\mathrm{bL}}$. This transconductance shows no charging lines associated with the right QD. Hence, for the applied gate voltages, the right QD is effectively decoupled from the leads. The typical resonant charge fluctuation time of the right QD is longer than the time a vertical sweep takes. In section (i), charging lines for the left QD and reconfiguration lines are still visible. Thus resonant charge exchanges via the left and the inter-dot barrier are fast compared to the duration of a vertical sweep. In section (ii), charging lines are more and more interrupted as $V_{\mathrm{bR}}$ is decreased. At the same time, the reconfiguration lines become elongated. This trend continues in section (iii) at even more negative $V_{\mathrm{bL}}$ where the charging lines are absent but the reconfiguration lines are vastly extended towards lower $V_{\mathrm{bR}}$.

While in figure~\ref{fig:Fig6b}(a) resonant charge fluctuations in the right QD are generally suppressed, between sections (i) to (iii), we clearly observe a transition to the situation in which charging events in both QDs become much slower than the vertical sweep time of about $t_{\mathrm{sweep}}=\mathrm{10~s}$. The elongated and clearly visible charge reconfiguration lines show that charge exchange between the two QDs is still fast compared to the sweep time. Hence in section (iii) of figure~\ref{fig:Fig6b}(a), the double QD tends to occupy a non-equilibrium charge configuration because of $\Gamma_{\mathrm{L}}\mathrm{, }\Gamma_{\mathrm{R}} \ll t_{\mathrm{sweep}}^{-1}$, while the charge distribution between the two QDs still tends to minimize the overall energy because of $\Gamma_{\mathrm{iD}} \gg t_{\mathrm{sweep}}$. From the length of the reconfiguration lines, we estimate for instance $\Gamma_{\mathrm{L}} \approx \mathrm{0.1~Hz}$ at $V_{\mathrm{bL}}=\mathrm{155~mV}$ and $V_{\mathrm{bR}}=\mathrm{175~mV}$, while $\Gamma_{\mathrm{L}} \gg \mathrm{1~Hz}$ if $V_{\mathrm{bL}}$ is increased only by about $\mathrm{20~mV}$. 

This measurement features two tendencies: tunneling rates are strongly susceptible to changes in gate voltage and tunneling rates are generally low. Both tendencies can be partly attributed to the large effective electron mass $m_{\mathrm{e}}^*$ in Si-based 2DES, since in first order, the tunneling rate across a barrier is proportional to $\exp(-\sqrt{m_{\mathrm{e}}^*\cdot E_{\mathrm{B}}})$, where the height of the tunnel barrier $E_{\mathrm{B}}$ is proportional to the applied gate voltages. 

The dependence of the inter-dot tunneling rate $\Gamma_{\mathrm{id}}$ on the applied gate voltage $V_{\mathrm{bR}}$ is more quantitatively investigated in figure~\ref{fig:Fig6b}(b). The data points are obtained by fitting the charging lines at the avoided crossings near the triple points marked by black arrows in figure~\ref{fig:Fig6a} for almost constant $V_{\mathrm{bL}} \approx \mathrm{203~mV}$. Exemplarily, two of the fits are shown as insets in figure~\ref{fig:Fig6b}(b). The distance between the solid lines is described by the function $\Delta E = \sqrt{(2\Delta)^2+(\hbar \Gamma_{\mathrm{id}})^2}+E_C$ where $2\Delta = (\mu_{\mathrm{R}}-\mu_{\mathrm{L}})$ is the asymmetry energy of the quantum-mechanical two-level system and $E_{\mathrm{C}}$ is the classical charging energy which represents the electrostatic coupling between the two QDs~\cite{VanderWiel2002,Huttel2005}. In order to fit the charging lines in a stability diagram based on applied gate voltages, in addition a linear transformation via the lever arms $\alpha_i^j$ (compare section~\ref{sec:p03_Transport}) and a rotation of the coordinate system is employed~\cite{Huttel2005}. Assuming $E_{\mathrm{C}}$ to be constant within a small range of applied gate voltages, we find best fits for $E_{\mathrm{C}} \approx \mathrm{435~\mu eV}$ and the tunneling rates $\Gamma_{\mathrm{id}}$ in figure~\ref{fig:Fig6b}(b). The solid line in figure~\ref{fig:Fig6b}(b) is a fit curve based on the WKB approximation for the inter-dot tunnel coupling $\Gamma_{\mathrm{id}} = \Gamma_{\mathrm{0}} \cdot \exp(- d \sqrt{m_{\mathrm{e}}^* E_{\mathrm{B}}}/\hbar) \sim \exp(\beta V_{\mathrm{bR}})$, where we assume for simplicity a constant width $d$ of the tunnel barrier and the barrier height $E_{\mathrm{B}}=E_{\mathrm{B}}^{\mathrm{0}}-\alpha_{\mathrm{B}} V_{\mathrm{bR}}$ and use $\alpha_{\mathrm{B}} V_{\mathrm{bR}}/E_{\mathrm{B}}^{\mathrm{0}} \ll 1$. The gate-barrier lever arm is defined by $\alpha_{\mathrm{B}}=E_{\mathrm{B}}/V_{\mathrm{bR}}$. Then the scaling factor $\beta$ depends on $m_{\mathrm{e}}^*$, $d$ and $\alpha_{\mathrm{B}}$. From the fitting procedure we find $\beta=\mathrm{0.056\pm0.023~mV^{-1}}$ which corresponds to $\Delta V_{\mathrm{bR}} \approx {40~mV}$ that are required to change the tunneling rate by one order of magnitude. This value is rather small compared to similar experiments with GaAs based double QDs \cite{Petta2004,DiCarlo2004}. The observed strong dependence of the tunneling rate on the gate voltage can in part be attributed to the higher effective electron mass in Si.

The tendency for small tunneling rates which strongly depend on gate voltages has been independently observed for QD-lead tunneling in figure~\ref{fig:Fig6b}(a) and inter-dot tunneling in figure~\ref{fig:Fig6b}(b). It has the following two direct implications: Due to the small tunneling rates, transport spectroscopy in the few-electron regime is more difficult in Si compared to double QDs defined in GaAs because of much smaller currents. On the contrary, the strong dependence of the tunneling rates on gate voltage is a chance for experiments which require time-dependent tunnel barriers - as often the case in quantum information processing.

\subsection{Pulsed Gate Experiments}
\label{sec:p05_PulsedGate}

Spin based quantum information processing requires fast initialization and manipulation of the spins in a double QD. We have combined charge sensing with pulsed gate operation~\cite{Petta2005} to demonstrate, as a first step, switching between two charge configurations. 

\begin{figure}
\begin{center}
\includegraphics[width=0.6\columnwidth]{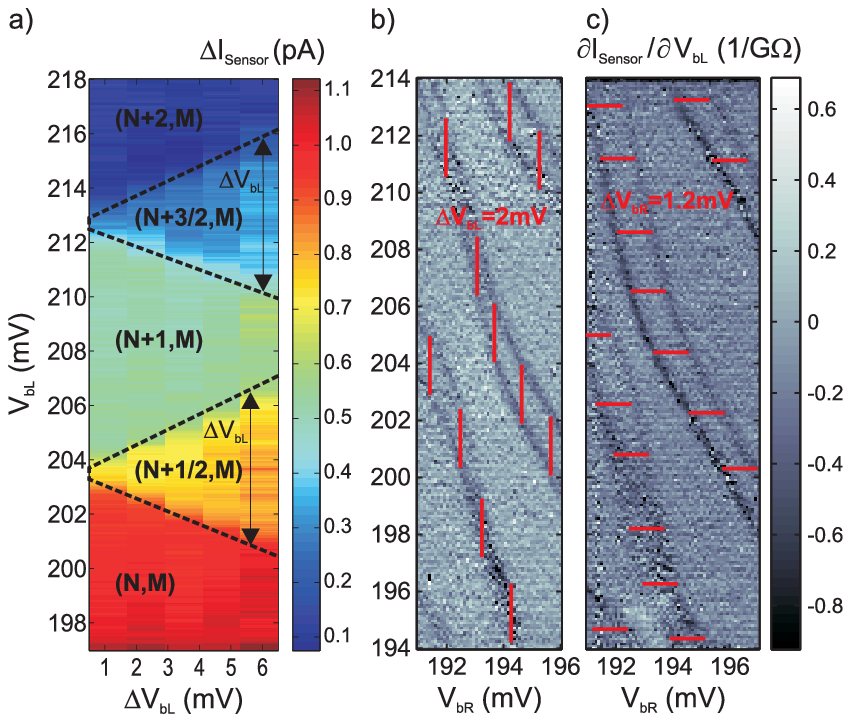}
\end{center}

\caption{a) Charge sensor current $\Delta I_{\mathrm{Sensor}} (V_{\mathrm{bL}})$ while a rectangular pulse sequence with 50~$\%$ duty cycle modulates $V_{\mathrm{bL}}$ by $\Delta V_{\mathrm{bL}}$. In between charge plateaus with integer occupation, intermediate plateaus emerge with increasing $\Delta V_{\mathrm{bL}}$. b) and c) Transconductance $\Delta \partial I_{\mathrm{Sensor}} / \partial V_{\mathrm{bL}}$ while pulsing with a 50~$\%$ duty cycle for the charge stability diagram shown in figure~\ref{fig:Fig5}(c). Pulses are applied to gate bL (b) or bR (c). Blue lines mark the pulse amplitude.}

\label{fig:Fig7}
\end{figure}
Figure~\ref{fig:Fig7}(a) shows the dc-charge sensor current $\Delta I_{\mathrm{Sensor}} (V_{\mathrm{bL}})$ for a fixed voltage $V_{\mathrm{bR}}$, while a rectangular pulse sequence of amplitude $\Delta V_{\mathrm{bL}}$ with a $50~\%$ duty cycle and a period of $\mathrm{1~ms}$ modulates $V_{\mathrm{bL}}$. With increasing pulse amplitude, two triangular-shaped intermediate plateaus with average charge configurations $(N+\mathrm{1/2},M)$ and $(N+\mathrm{3/2},M)$ develop. The vertical extension  of the intermediate plateau is identical to $\Delta V_{\mathrm{bL}}$. This result demonstrates the effect of a sudden change in a gate voltage, namely a shift of the local potential which defines the double QD. This local potential difference results in a shift of the entire stability diagram for the duration of the pulse. For the applied pulse train with a 50~$\%$ duty cycle, we therefore expect to find two copies of the stability diagram shifted according to the pulse direction and amplitude. This can be seen in figure~\ref{fig:Fig7}(b) and (c) where the pulses were applied to gate bL (b) and bR (c) with amplitudes of $\Delta V_{\mathrm{bL}} = \mathrm{2~mV}$ and $\Delta V_{\mathrm{bR}} = \mathrm{1.2~mV}$, respectively. Clearly, the charging lines split into doublets of parallel charging lines with the corresponding distance $\Delta V_{\mathrm{bL}}$ or $\Delta V_{\mathrm{bR}}$. Due to the strong inter-dot coupling, the reconfiguration lines are rather broadened than split. 

The low QD-lead tunneling rates $\Gamma_{\mathrm{L}}$ and $\Gamma_{\mathrm{R}}$ in the few-electron regime restricts our pulse repetition rates to no more than about $\mathrm{10~kHz}$. We have also performed pulse repetition rates up to approximately $\mathrm{5~MHz}$ limited by our instruments in the regime of larger QD-lead tunneling rates.

\section{Conclusion}
\label{sec:p06_Conclusion}

In summary, we have performed direct transport spectroscopy through a few-electron Si/SiGe double QD, charge-sensing with a remote single QD sensor and pulsed-gate measurements.  We deduce material-specific implications for the implementation of double QDs and spin qubits.
An important parameter influencing the transport properties of our QD devices is the comparatively large effective electron mass $m_{\mathrm{e}}^*$ in Si-based 2DES. It enhances the dependence of tunneling rates on gate voltage and correspondingly can cause overall low tunneling rates across electrostatic barriers. Additionally, the large $m_{\mathrm{e}}^*$ contributes to a small Fermi energy, together with the two-fold valley degeneracy. The combination of low tunneling rates and small Fermi-energies hampers linear response transport spectroscopy with a current flowing across a double QDs in the few-electron regime. However, these difficulties can be circumvented by smaller feature sizes in future devices. From another perspective, the relatively strong scaling of tunneling rates with gate voltage can be exploited to implement efficient tuning of tunneling rates by pulsing gate voltages with a limited amplitude.
As an alternative to transport spectroscopy, a spin qubit can also be operated at a constant overall charge of a double QD in combination with charge spectroscopy. Based on such measurements, we find QDs in our Si/SiGe heterostructure devices still exposed to more charge noise than mature GaAs-based devices. Yet, our experiments also demonstrate a promising tendency towards quiet operation of the double QD, when manipulating gate voltages only in a limited range. These results suggest a realistic path towards Si-based quantum information processing.

The key advantage for Si-based qubits is the reduced interaction of confined electron spins in Si with their volatile crystal environment that gives rise to a number of decoherence mechanisms. Phonon-mediated back-action of a remote charge sensor on a qubit, which has been observed in GaAs based QDs~\cite{Schinner2009,Harbusch2010}, can be expected to be much weaker in Si. Indeed, the electron-phonon coupling is reduced (\emph{e.g.} no piezo-electricity) and the low Fermi energy reduces the band-width for phonon-mediated interaction~\cite{Schinner2009}. Furthermore, the spin-orbit coupling is weak and the hyperfine interaction in natural Si crystals is reduced compared to GaAs. Most importantly, our results show that the presented device layout with the possibility of almost zero Overhauser field in recently realized isotopically purified $^{\mathrm{28}}$Si 2DES~\cite{Sailer2009} makes Si-based QDs a promising candidate for spin qubits with coherence times much larger than those that can be realized in GaAs/AlGaAs heterostructures.

\begin{acknowledgments}

Financial support by the Deutsche Forschungsgemeinschaft via SFB 631 and the "Nano Initiative Munich (NIM)" is gratefully acknowledged. We thank Daniela Taubert, Daniel Harbusch and Stephan Manus for helpful discussions.

\end{acknowledgments}

\end{document}